\documentstyle[editedvolume]{crckapb}
\input{psfig}

\newcommand{\sigo}{\sigma_{\rm o}}

\newcommand{\sigeight}{\sigma_{8}}
\newcommand{\om}{\Omega_{0}}
\newcommand{\lx}{L_{X}}
\newcommand{\tx}{T_{X}}

\begin{opening}
\title{High Redshift X-ray Clusters and $\om$ }
\author{R. SADAT}
\institute{Affiliation\\
Observatoire astronomique de Strasbourg, 11 rue de l'universit\'e. 67000 Strasbourg, France.\\
and\\
I.A.P. 98bis Bd. Arago, 75014 Paris, France}
\end{opening}
\begin{document}
\vspace{-5mm}
\section{Introduction}
\vspace{-1.5mm}
	Clusters of galaxies are undoubtedly the largest virialized systems in the universe, they offer a powerful tool in constraining cosmological parameters. For example, their present-day abundance has been used to constrain parameters such as the {\it shape} $n$ and the {\it amplitude} ${\sigeight}$ of the mass fluctuations as well the cosmological density parameter ${\om}$. However this constraint is degenerate in ${\sigeight}$ and ${\om}$. Recently, Oukbir \& Blanchard (1997; hereafter OB97), have nicely shown that the {\it evolution} of X-ray temperature distribution function {\it does depend strongly} on ${\om}$, and claimed that the combined knowledge of the abundance of X-ray clusters together with the evolution of the luminosity-temperature relation would allow us to strongly constrain the cosmological density parameter independently of the power spectrum. I will present the result of our first attempt to apply this test to current X-ray data (Sadat et al. 1997). \vspace*{-1.5mm} 

\section{Basic Recipe}

The Press-Schechter (1974) (PS) simple analytical formalism allows us to calculate the comoving number density of dark matter halos of a given mass $M
$. This approximation of the {\it mass function} seems to fit remarkably the simulations and then can be applied to clusters with present--day X-ray temperatures of $\sim 1-10\;$ keV. For more details on the derivation of PS formula see Bartlett 1997. To relate the {\it mass} to observable quantities, we prefer to deal with X-ray observations than with optical data because of the projection and contamination effects which may affect optical cluster observations. In principle, $\lx$ is the simplest quantity to measure, nevertheless one must avoid to relate the {\it mass} to $\lx$, because of the strong dependence of the gas emissivity on the spatial distribution of the gas in the core of the cluster which physics is not well understood. In contrast, X-ray temperature--mass relation is better understood and follows simple physics (${\tx}\propto M/R$). The X-ray gas temperature is fairly well predicted by hydrodynamical simulations (Evrard et al. 1996). These authors have shown the existence of a tight relation between $\tx$ and the mass: $T_{X}=(6.8h^{2/3} keV)M^{2/3}(1.+z)$. Thus, it is possible to use this simple relation to derive the expected X-ray gas {\it temperature} distribution function ${\phi(T)}$ from the {\it mass} function and then compare it to observations. Unfortunately, this {\it temperature} function is only available at $z=$ 0, we do not yet have any information on the evolution of ${\phi(T)}$ with $z$.\vspace*{-1.5mm} 

\subsection{Evolution of X-ray clusters: a test for ${\om}$?}
	Now that we have built a self-consistent modelling of X-ray clusters, one may address the following question: what one would expect from high-z X-ray clusters?
Under some reasonable simplifications, the {\it mass function} can be written as (Blanchard et al. 1992):\vspace{-0.4mm}
\begin{equation}
N(M, z) = N(M, z = 0) \frac{F\left(\nu_0A(z)\right)}{F(\nu_0)}A(z).
\end{equation}
and \vspace{-0.4mm}$
N(M, z=0) = \frac{{\overline \rho}}{M^2} \nu_0
\frac{d\log \sigo}{d\log M}F(\nu_0).
$
where
${\nu_0}$ being $\delta_{c,0}(z=0)/\sigo(M)$. \\
F is well fitted by an ${\rm exp}(-\nu^{2}/2)$ independently of the power spectrum $P(k)$ (from numerical  simulations) and A(${\om}$,z) is the growing rate. Equation (2) demonstrates clearly that the rate of cluster {\it evolution} with redshift $z$ is mainly driven by the density parameter $\om$, independently of the power spectrum index and ${\sigeight}$. In open universe, we should expect more clusters at high $z$ than in a critical universe. Consequently, the study of the {\it evolution} of the cluster number density, once normalized to present--day abundance, would strongly constrain $\om$. However, the observation of such information that is -- the {\it evolution} of the X-ray temperature distribution function-- is far from being reached.
The EMSS cluster redshift survey is up to now the largest sample of well-controlled X-ray selected clusters. This has been studied in detail by OB97 who found that in order to achieve a self-consistent modeling of the X-ray data, one needs to introduce a  negative evolution in the ${\lx}$--${\tx}$ relation:
 $L_{X} \propto T_{X}^3(1+z)^{\beta}$ in order to match the data in an open universe (with ${\beta}$=-2.3), otherwise, i.e in the case of {\it no evolution} low ${\om}$ models are simply ruled out. A recent analysis based on published clusters abundance has led to the same conclusion (Blanchard \& Bartlett 1998, BB98) and Fig. 1. Therefore, the evolution of the ${\lx}$--${\tx}$ relation would allow us to disentangle between open and critical universe. Moreover, by iterating this analysis for different ${\om}$, Sadat et al. (1997) have shown that the best fitting ${\beta}$ is tightly related to ${\om}$ accordingly to ${\beta}=4. \times {\om}$ - 3. \vspace*{-1.5mm} 
\begin{figure}[t]
\centerline{\psfig {figure=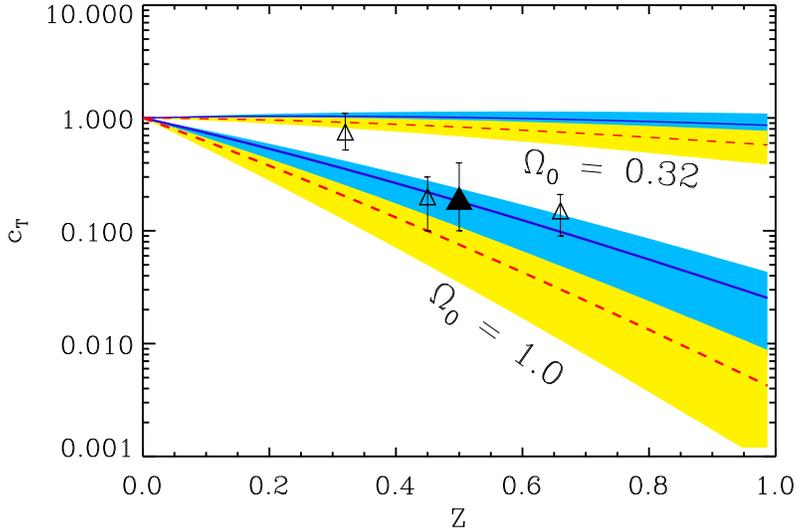,height=8cm,angle=0}}{\vspace{-6mm}}
\caption{\small \sl Evolution with $z$ of the relative number of clusters at a given ${\tx}$ (from BB98). The solid and dashed lines correspond respectively to 4 keV and 6 keV clusters. Filled and (empty) triangles are 4 keV and (6 keV) clusters.\vspace{-3mm}}
\section{Application of the test}

	Our sample results from a compilation of 57 high-z clusters among which $\sim$ 30 clusters at $z\ge$ 0.26. Local observations have also been included. We added few clusters for which accurate mass estimate is available. We have derived the temperature, using simple scaling relations. In order to apply the test discussed above, that is to measurement of the evolution in ${\lx}$--${\tx}$, we have fitted the following power law $ C(z) = {\alpha} (1+z)^{\beta}$ to 
observational  coefficient $C_i={L^{bol}_i}/{0.05 T_i^3}$. The result is presented in Fig. 2. While ${\alpha}= 1$ as expected, the best ${\beta}$ value ranges in [0.--1.0] interval. This result shows no evidence of a significant evolution in the ${\lx}$--${\tx}$ relation in agreement with previous investigations.\vspace*{-1.5mm} 

\end{figure}
\begin{figure}[h]
\centerline{\psfig {figure=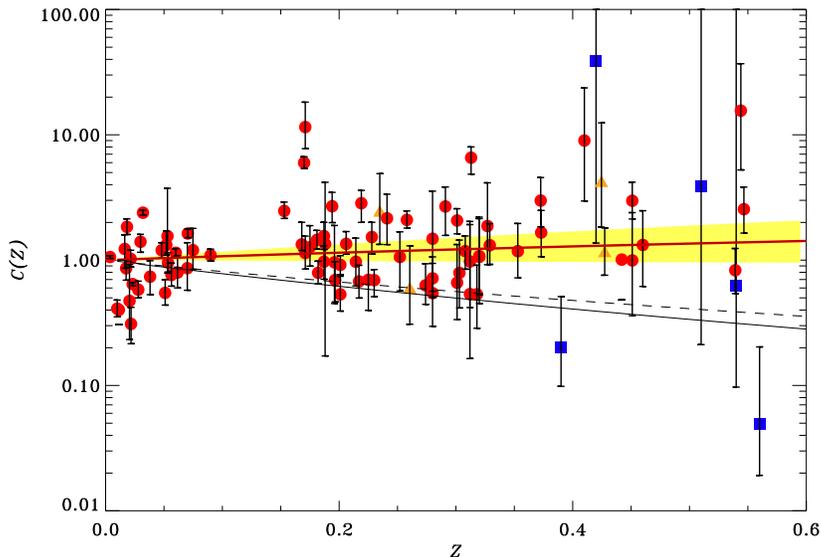,height=8cm,angle=0}}{\vspace{-6mm}}
\caption{\small \sl Coefficient $C_i$ versus $z$ from Sadat et al. (1997). The error bars are $1 \sigma$. The thick line represents the best fitting power--law, the shaded area represents an estimate of the 90\% confidence range. The prediction for an open universe is represented by the thin line. \vspace{-3mm}}
\end{figure}

\subsection{What do the data tell us on ${\om}$ ?}
	As discussed in section 2.1, in low ${\om}$ universe, a strong evolution is required in order to match the EMSS observations. Therefore, current data which show no significant evolution suggest that ${\om}$$>$0.2. Now, we can go further and give an estimation of ${\om}$ from the best fit value of ${\beta}$. Using our simple relation between ${\beta}$ and ${\om}$, one can directly derive the value of ${\om}$ corresponding to the observed ${\lx}$--${\tx}$ evolution rate ${\beta}$: we find ${\om}$=0.85 ${\pm}$0.2, although a full analysis of the errors still remains to be done. \vspace*{-1.5mm}

\section{Conclusions}
	The evolution of the cluster temperature distribution function is crucial for cosmology, it represents a powerful test for ${\om}$. The evolution of the ${\lx}$--${\tx}$ relation provides us with a useful variant of this test (OB97). We have reported the first attempt to apply this test to observations and found that --i) current X-ray data show no evidence of a strong evolution in the ${\lx}$--${\tx}$ relation,  --ii) the lack of evolution is consistent with high ${\om}$ universe, --iii) the evolution of the ${\lx}$--${\tx}$ relation has been quantified and suggests that $0.75 <$ ${\om}$ $ < 1$.\end{document}